# Emergent toroidal induction in a polar Weyl ferromagnet

Yuuri Suzuki[1*], Yukako Fujishiro[1,2,3,4*†], Masataka Mogi[1], Juba Bouaziz[5,6], Takahiro Anan[1], Akiko Kikkawa[3], Daiki Yamaguchi[1,3], Max T. Birch[3], Yuto Kiyonaga[1], Minoru Kawamura[3], Yasujiro Taguchi[3], Takahiro Morimoto[1], Naoto Nagaosa[3,7], Ryotaro Arita[3,5] and Yoshinori Tokura[1,3,8†]

*These authors contributed equally to this work.*

†*Corresponding authors.*

[1] *Department of Applied Physics, The University of Tokyo, Tokyo 113-8654, Japan*

[2] *Institute of Engineering Innovation, The University of Tokyo, Tokyo, 113-8656, Japan*

[3] *RIKEN Center for Emergent Matter Science (CEMS), Saitama 351-0198, Japan*

[4] *RIKEN Pioneering Research Institute (PRI), Saitama 351-0198, Japan*

[5] *Department of Physics, The University of Tokyo, Tokyo 113-8654, Japan*

[6] *Faculty of Physics, University of Duisburg-Essen, Duisburg 47057, Germany*

[7] *Fundamental Quantum Science Program (FQSP), TRIP Headquarters, RIKEN, Wako 351-0198, Japan*

[8] *Tokyo College, The University of Tokyo, Tokyo 113-8654, Japan*

**Spin–orbit coupling (SOC) underpins modern spintronics by enabling the electrical generation of spin torques. Its reciprocal counterpart, where magnetization dynamics produce electromotive forces via spin-dependent Berry phase, is known as emergent electromagnetic induction (EEMI). However, this effect has been observed only in magnetic textures with spatial gradients such as domain walls, helices, and skyrmions. Here we demonstrate that even a uniform ferromagnet can host this effect through a previously unrecognized Berry-phase mechanism inherent to non-centrosymmetric conductors. In the polar Weyl ferromagnet PrAlGe, an applied alternating current generates spin-orbit torques that drive collective magnetization dynamics; the resulting emergent toroidal moment $\boldsymbol{T} = \boldsymbol{P} \times \boldsymbol{M}$ (with $\boldsymbol{P}$ the crystal's polar axis and $\boldsymbol{M}$ the net magnetization) acts as a gauge potential, whose time-derivative $\mathrm{d}\boldsymbol{T}/\mathrm{d}t$ induces Hall voltages. This contribution appears specifically in the out-of-phase component of the AC Hall response and scales linearly with frequency, providing direct evidence for EEMI. First-principles calculations further reveal that this toroidal vector encodes the collective motion of Weyl nodes in momentum space. These findings establish "emergent toroidal induction" as a new manifestation of spin–orbit entanglement, unifying Berry phase, topology, and spin dynamics, and opening a pathway toward intrinsic, energy-efficient spin–charge interconversion.**

The interconversion between electricity and magnetism lies at the heart of modern electronics, governing power generation and information processing. Faraday's law of induction states that a temporally changing magnetic flux induces an electric field. Recent advances in condensed matter physics have uncovered a quantum analogue of this principle[1]: the time evolution of electron spins can likewise induce electric responses, not through changes of magnetic flux, but through geometric phases generated in spin space. This "emergent electromagnetic induction" (EEMI) has been demonstrated in non-uniform magnets, where moving domain walls[2–4], current-driven skyrmions[5,6] and short-pitch spin helices convert spin dynamics into measurable voltages via spin-motive forces[7–9] (Fig. 1**a**).

A key common feature in these realizations is their reliance on spatially textured magnetization, whose gradients act as effective vector potential for the electrons and whose time dependence produces emergent electric fields. While this EEMI originates from a Berry phase accumulated in spin space, theory suggests that even a spatially uniform ferromagnet can host an analogous effect when spin-orbit coupling (SOC) and inversion symmetry breaking are present[10]. In this regime, SOC generates a spin-dependent Berry gauge field that links electron motion to the instantaneous orientation of a homogeneous magnetization[11].

A convenient starting point is the SOC Hamiltonian,

$$H_{\mathrm{SOC}} \propto (\nabla V \times \boldsymbol{p}) \cdot \boldsymbol{S} \tag{1}$$

where $\boldsymbol{p}$ and $\boldsymbol{S}$ are the electron momentum and spin, and $\nabla V$ is the gradient of the Coulomb potential. In a polar crystal with a polarity $P$ and net magnetization $M$, where the conduction-electron spin adiabatically follows the localized moments, these microscopic fields can be replaced by their macroscopic averages, $\langle -\nabla V \rangle \propto \boldsymbol{P}$ and $\langle \boldsymbol{S} \rangle \propto \boldsymbol{M}$, leading to

$$(\nabla V \times \boldsymbol{p}) \cdot \boldsymbol{S} \simeq (\langle -\nabla V \rangle \times \langle \boldsymbol{S} \rangle) \cdot \boldsymbol{p} \propto (\boldsymbol{P} \times \boldsymbol{M}) \cdot \boldsymbol{p} \tag{2}$$

The combination $\boldsymbol{T} = \boldsymbol{P} \times \boldsymbol{M}$ is the toroidal moment, so that Eq (2) can be written schematically as $H_{\mathrm{SOC}} \propto \boldsymbol{T} \cdot \boldsymbol{p}$. Note that in magnetoelectric multiferroics[12,13], the toroidal moment is known to serve as a key parameter capturing the intertwined nature of electricity and magnetism. Crucially, a term linear in momentum, $\boldsymbol{T}$ acts as a uniform effective gauge potential. For kinetic electrons $H_0(\boldsymbol{p}) = \boldsymbol{p}^2/2m^*$ ($m^*$: effective electron mass), the SOC-induced perturbation is equivalent to a momentum shift:

$$H(\boldsymbol{p}) = H_0(\boldsymbol{p}) + \boldsymbol{T} \cdot \boldsymbol{p} \cong H_0(\boldsymbol{p} - \boldsymbol{a}_{\mathrm{eff}}), \tag{3}$$

with $\boldsymbol{a}_{\mathrm{eff}} \propto \boldsymbol{T}$. As a result, the entire dispersion is displaced in momentum space, and the toroidal moment plays the role of an effective vector potential. The emergence of the vector potential even in the uniform magnetic field essentially comes from the fact that the uniform magnetization with SOC can be mapped to a nontrivial spin texture with a suitable unitary transformation, where itinerant electrons experience nontrivial real-space geometry from the background magnetization.

When the magnetization $M(t)$ varies in time, the associated toroidal moment $\boldsymbol{T}(t)$ produces a time-dependent effective gauge field. Electrons therefore experience an emergent electric field,

$$\boldsymbol{e}_{\mathrm{em}} = -\frac{\partial \boldsymbol{a}_{\mathrm{eff}}}{\partial t} \propto -\frac{\partial \boldsymbol{T}}{\partial t}, \tag{4}$$

driven by the dynamics of the toroidal moment. Realizing such "toroidal induction" requires a materials platform combining ferromagnetism, strong SOC, and a polar axis so that $\boldsymbol{T} = \boldsymbol{P} \times \boldsymbol{M}$ is defined.

To drive a time-dependent $\boldsymbol{T}(t)$, various stimuli can be used, including magnetic fields or magnetic resonance. Here we focus on current-induced spin dynamics mediated by spin orbit torques (SOTs)[14]. In non-centrosymmetric conductors, SOC produces spin-split bands with spin-momentum locking[15,16]. When a current flows, the resulting nonequilibrium spin accumulation exerts field-like and damping-like torques on the magnetization. In a polar ferromagnet whose magnetic easy axis aligns with the polar axis, the equilibrium toroidal moment vanishes ( $\boldsymbol{T} = \boldsymbol{P} \times \boldsymbol{M} = 0$ ), but current-driven magnetization oscillations $\delta\boldsymbol{M}(t)$ generate a time-dependent toroidal moment $\boldsymbol{T}(t) = \boldsymbol{P} \times [\boldsymbol{M} + \delta\boldsymbol{M}(t)] = \boldsymbol{P} \times \delta\boldsymbol{M}(t)$, and thus a dynamically oscillating emergent electric field (Fig. 1**b**).

The polar Weyl ferromagnet PrAlGe, a member of the non-centrosymmetric $I4_1md$ family *R*AlSi/*R*AlGe (*R* = rare earth)[17–20], satisfies these conditions (Fig. 1**c**). Its crystal structure enforces a built-in SOC field, while ferromagnetic order along the polar *z*-axis breaks time-reversal symmetry. Below the magnetic transition temperature $T_{\mathrm{C}} = 15$ K (see Supplementary Fig. 1), PrAlGe exhibits nearly Ising-type magnetization[21] (Fig. 1**d**) and a large anomalous Hall conductivity (Fig. 1**e**), reflecting strong coupling between magnetism and Weyl-band Berry curvature[22–24].

In this work, we demonstrate that an AC current in PrAlGe drives SOT-induced magnetization tilting, and that the dynamics of these spin–momentum–locked bands generate a spin-dependent vector potential whose temporal oscillation acts as a source of electromotive force. Experimentally, we detect this as a Hall voltage whose imaginary component scales linearly with frequency, providing direct evidence of EEMI arising from the SOC-induced gauge potential in a system with net magnetization. From a phenomenological viewpoint, this response corresponds to $\partial \boldsymbol{T}/\partial t$, whereas first principles calculations reveal its microscopic origin in SOC-mediated coupling between spin dynamics and Weyl band topology, leading to a collective displacement of Weyl nodes (i.e., monopoles and anti-monopoles in momentum space). Together these perspectives establish emergent toroidal induction as a unified manifestation of topology, magnetism, and spin dynamics in a magnetic quantum material.

**Current-induced magnetization dynamics**

To probe the current-induced magnetization dynamics, we fabricated micro-structured Hall bars from single-crystalline PrAlGe using focused-ion-beam milling, enabling current densities exceeding $10^8\ \mathrm{A/m^2}$. The sample temperature for each current value was adjusted so that the longitudinal resistance matched its low-current value at 11.7 K, effectively compensating Joule-heating (see Supplementary Fig. 2). When a direct current $j_{\mathrm{DC}}$ is applied along the *x*-axis under a magnetic field along *z*, the anomalous Hall conductivity $\sigma_{xy}^{\mathrm{A}}$ decreases systematically with increasing current (Fig. 2**a**). This reduction indicates a tilt of the magnetization away from the easy *z*-axis, *i.e.*, a decrease in $M_z$ due to current-induced SOT. Although the detailed direction of SOT cannot be determined in this DC measurement, the magnetization tilt is corroborated by our non-linear Hall effect measurements[25,26].

The reduction of $\sigma_{xy}^{\mathrm{A}}$ scales linearly with current density, reaching ~10 % at $j_{\mathrm{DC}} = 8 \times 10^8\ \mathrm{A/m^2}$ (Fig. 2**b**). To quantify the tilt angle, we compared the normalized change in $\sigma_{xy}^{\mathrm{A}}$ under current with that produced by rotating the external magnetic field. For instance, as shown in Extended Data Fig. 1**d**, the reduction of anomalous Hall conductivity by applying $j_{\mathrm{DC}} = 6 \times 10^8\ \mathrm{A/m^2}$ corresponds to the case for the tilt angle $\theta_M = 8.8^{\circ}$ (see Methods for detail). The resulting magnetization angle $\theta_M$ (Fig. 2**c**) increases linearly with $j_{\mathrm{DC}}$, reaching approximately $\sim 10^{\circ}$ at $6.5 \times 10^8\ \mathrm{A/m^2}$.

Notably, the observed ~10 % suppression of $\sigma_{xy}^{\mathrm{A}}$ far exceeds the geometric expectation of only 1.5 % from the reduction of $M_z$ $\left(\Delta\sigma_{xy}^{A} \propto \sigma_{xy}^{A}(1-\cos 10^{\circ})\right)$ (Extended Data Fig. 1**d**). This discrepancy suggests that current-induced canting causes substantial modifications to the electronic structure rather than a simple reduction of magnetization magnitude. Since the anomalous Hall effect is highly sensitive to Berry curvature near the Fermi level, the large suppression of $\sigma_{xy}^{\mathrm{A}}$ signals a dramatic redistribution of Berry curvature triggered by spin–orbit–mediated coupling between spin orientation and band topology.

**Weyl-node dynamics and Berry curvature reconstruction**

To explore the effect of magnetization tilt on the electronic structure, we performed density-functional theory (DFT) calculations for magnetization tilt angles $\theta_M = 0°$ and $15°$ along the $x$-direction. Figure 2**d** shows the calculated band structures (top) and the corresponding change in $z$-component of Berry curvature ($\Delta\Omega_z(\boldsymbol{k})$) (bottom) showing a pronounced redistribution (see Extended Data Fig. 2 for BZ definition).

While the simple shift of the Weyl points does not produce the net effect when integrated over the momentum space[27], we emphasize that the contributions to EEMI from a pair of Weyl nodes do not cancel out by magnetization tilt in Rashba-Weyl bands. The effective Hamiltonian of a SOC–induced Weyl semimetal with in-plane Rashba-SOC is typically given by $\lambda(k_y\sigma_x - k_x\sigma_y) + \eta v_z\sigma_z$ for Weyl ($\eta = +1$) and anti-Weyl ($\eta =$ $-1$) nodes, where $\lambda$ is the SOC strength, and $v_z$ is the Fermi velocity. This model contains the built-in gauge field where the effective vector potential along the *y*-direction generated by the magnetization dynamics along $x$-direction, $\delta m_x \sigma_x$, takes the same value $\delta m_x / \lambda$ for both Weyl and anti-Weyl nodes, whose contributions add up. Therefore, it is also informative to focus on the Weyl-node dynamics by tracing their positions. Indeed, we found that the modest magnetization tilt of $\theta_M = 15°$ is sufficient to induce a dramatic reconfiguration of Weyl nodes (Fig. 2**e**), including substantial shifts of existing nodes as well as the emergence of new ones. Figure 2**f** shows an enlarged view of the area highlighted by the dashed square in Fig. 2**e**, where the displacement of Weyl nodes is clearly resolved: the effective shift of the Weyl nodes is most evident along the $k_y$-direction. Furthermore, two-dimensional maps of $\Omega_z(\boldsymbol{k})$ (Fig. 2**g**) further reveal that the Berry curvature distribution becomes strongly asymmetric under tilt, accompanied by the appearance of new singular features near the Fermi level. The prominent change of Berry curvature is highlighted by dotted circles (Fig. 2**g**). It is noteworthy that while the change of Berry curvature along the $k_x$-direction is rather symmetric with respect to $k_x = 0$, that of the $k_y$-direction exhibits a pronounced antisymmetric change with respect to $k_y = 0$. This clearly indicates that the tilt of magnetization along the *x*-direction induces asymmetric quantum geometry in the y-direction. As discussed earlier, the over-all shift of the electronic band structure, including the vector potential change in momentum space by Weyl node dynamics (Fig. 2**h**), is encoded within the change of the toroidal moment, which leads to the emergent electric field.

**Emergent toroidal induction**

To directly probe the emergent electric field ($\boldsymbol{e}$), we measured the alternating-current response associated with time-varying magnetization dynamics. Starting from a state where the magnetization is aligned with the easy axis ($\boldsymbol{M} \parallel \boldsymbol{P}$), an alternating current $I = I_0 e^{-i\omega t}$ induces an in-plane magnetization component $\delta\boldsymbol{M}$ through the SOT.

$$\boldsymbol{e} \propto -\frac{\boldsymbol{P}\times\delta\boldsymbol{M}(I)}{\delta t} \propto -\frac{\partial I}{\partial t} \propto i\omega e^{-i\omega t}. \quad (4)$$

Consequently, the emergent electric field driven by current-induced magnetization dynamics manifests as the imaginary component of complex impedance, $\mathrm{Im}\, Z^{\omega}$. We applied an alternating current along the $x$-direction and compared the magnitude of the longitudinal ($\mathrm{Im}\, Z^{\omega}_{xx}$) and transverse ($\mathrm{Im}\, Z^{\omega}_{yx}$) components using the quality factor, defined as the ratio of the imaginary to real parts of the complex impedance. As shown in Fig. 3**a**, $\mathrm{Im}\, Z^{\omega}_{yx}/\mathrm{Re} Z^{\omega}_{yx}$ dominates over $\mathrm{Im}\, Z^{\omega}_{xx}/\mathrm{Re} Z^{\omega}_{xx}$. We thus focus on the transverse EEMI component ($\mathrm{Im}\, Z^{\omega}_{yx}$) in the following analysis. Note that any extrinsic phase rotation originating from the measurement circuit was corrected by referencing the data taken at 30 K above the Curie temperature ($T_c$) (see Methods).

Figure 3**b** shows the magnetic-field dependence of $\mathrm{Im}\, Z^{\omega}_{yx}$ for various current densities $j_{\mathrm{AC}}$ –which decreases with increasing magnetic field–consistent with the suppression of magnetization tilting at higher fields. The inset of Fig. 3**b** displays the current-density dependence, revealing that $\mathrm{Im}\, Z^{\omega}_{yx}$ increases with $j$, indicating a nonlinear response. This nonlinearity likely arises from current-induced changes in the electronic structure, as evidenced by the non-trivial dependence of the anomalous Hall effect on the magnetization tilt angle (Extended Data Fig. 1**d**). Although the anomalous Hall effect is not directly linked to the emergent electric field, its pronounced angle dependence supports the interpretation that the observed emergent inductance becomes

nonlinear through current-driven modifications of the electronic structure. Figure 3**c** presents $\mathrm{Im}Z_{yx}^{\omega}$ as a function of magnetic field at various temperatures; the signal weakens with increasing temperature, consistent with the reduction of magnetization. Finally, Figure 3**d** shows the temperature dependence of $\mathrm{Im}Z_{yx}^{\omega}$, which drops sharply above $T_c$. This behavior demonstrates that the observed emergent electric field is an intrinsic consequence of magnetization dynamics. Notably, a recently proposed Joule-heating-based model of EEMI[28,29] is ruled out by our higher-harmonic transport measurements (see Supplementary Fig. 5).

**Frequency dependence of EEMI**

Figure 4**a** shows the frequency dependence of $\mathrm{Im}Z_{yx}^{\omega}$ at various magnetic fields. For each magnetic field, $\mathrm{Im}Z_{yx}^{\omega}$ increases linearly with frequency up to approximately $f \sim 2$ kHz, consistent with the behavior predicted by equation (4). As shown in Fig. 4**b**, the inductance $L_{yx} = \mathrm{Im}Z_{yx}^{\omega}/2\pi f$ remains constant at low frequencies, followed by Debye-type relaxation at higher frequencies. The microscopic origin of this relaxation remains an open question. Figure 4**c** presents the magnetic-field dependence of $\mathrm{Im}Z_{yx}^{\omega}$ at several frequencies, all showing a systematic suppression with increasing magnetic field. This behavior further supports the interpretation that $\mathrm{Im}Z_{yx}^{\omega}$ originates from magnetization dynamics, which are progressively suppressed under stronger fields. Together, these results confirm that the observed emergent electric field reflects the time derivative of magnetization-driven effective vector potential, establishing a clear correspondence between experimental frequency dependence and theoretical framework of EEMI.

**Discussion**

Emergent electric fields driven by magnetization dynamics have recently been reported in helimagnets[8,9], domain-wall systems[30], and skyrmion lattices[6,31], typically appearing as longitudinal impedance or nonlinear resistance. Our work reveals a qualitatively distinct regime: an emergent electric field in a homogeneous bulk metal with net magnetization, detected through the imaginary Hall response. This provides the direct evidence that the induction originates from a time-dependent SOC-induced vector potential in a system with net magnetization, rather than from real-space spin textures or topological defects that have dominated previous studies. In a microscopic point of view, the result can be regarded as a long-anticipated mechanism in which the motion of Weyl points acts as an intrinsic electromotive source.

The magnetization dynamics in PrAlGe are activated by a remarkably low current density (~$10^8$ A m$^{-2}$), in sharp contrast with the thresholds of $10^{10}$–$10^{11}$ A m$^{-2}$ typically required in spintronic thin-film heterostructures. A similarly low threshold in CeAlSi identified via second-harmonic generation[32] suggests that bulk polar magnets in the RAlX (*X*=Si or Ge) family may provide an unusually efficient route to current-driven magnetic manipulation. Full magnetization reversal is not achieved in PrAlGe, largely because the low magnetic transition temperature limits the maximum current-density allowed in the experiment when Joule-heating is considered, while the unexpectedly small coercivity and formation of multidomain states hinder field-free switching. These observations indicate a clear materials-design direction: polar magnetic systems with higher coercivity and elevated magnetic transition temperatures could realize field-free, full magnetization switching accompanied by substantially enhanced emergent electric fields tied to topological band reconstruction.

Beyond materials optimization, our results uncover an intrinsic spin–charge conversion mechanism rooted in the topological dynamics of Rashba-Weyl band structures, providing a complementary paradigm to interfacial spin–orbit torques. The linear-frequency imaginary Hall impedance represents a condensed-matter analogue of a classical inductor, where the magnetization-driven vector potential plays the role of the flux penetrating the coil. Harnessing such topological inductive responses may ultimately enable all-metallic, energy-efficient elements for ultracompact spintronic architectures[33].

## Methods

### Sample preparation and device fabrication

Polycrystals of PrAlGe were synthesized by arc melting in argon atmosphere. The polycrystal samples were characterized by powder *x*-ray diffraction (XRD). The obtained polycrystals were shaped into rods using the arc melting method. The single crystal of PrAlGe was grown under argon gas flow by a floating zone (FZ) method from the rods. The single crystal was characterized by magnetization.

Devices were fabricated from the single crystals using a Hitachi NB5000 focused ion beam (FIB) system equipped with a Ga ion source. $SiO_2$/Si substrates were cleaned and patterned with Ti/Au (50 nm/1950 nm) electrodes using UV photolithography. Electrical contacts between the Au pads and the PrAlGe slab were formed by W deposition using the NB5000 system. The final Hall-bar geometry was defined by additional Ga FIB milling. The dimensions for FIB-fabricated devices are 20 μm × 6 μm × 1.2 μm for device#2, 20 μm × 6.4 μm × 1.2 μm for device#4, 2 0 μm × 5.9 μm × 1.2 μm for device#N1.

### Magnetization and transport measurements

Magnetization measurements were carried out using a magnetic property measurement system (MPMS, Quantum Design). The single-crystal sample was mounted at the end of a plastic straw using varnish and Kapton tape. Transport measurements were carried out using a Quantum Design Physical Property Measurement System (PPMS) with a standard four-probe configuration. Direct-current transport was measured using the built-in resistivity option. Longitudinal and Hall resistivities were obtained by symmetrizing and anti-symmetrizing the signals with respect to magnetic field, respectively. The anomalous Hall conductivity $\sigma_{xy}^A$ was evaluated as follows: we assumed $\rho_{yx} = R_0 B + \rho_{yx}^A$, where $R_0$ is the ordinary Hall coefficient and $\rho_{yx}^A$ is the anomalous Hall resistivity. $R_0 B$ was obtained by fitting the high-field region of $\rho_{yx}$ with a linear function, and $\rho_{yx}^A$ was extracted by subtracting $R_0 B$ from the total $\rho_{yx}$. Then, $\sigma_{xy}^A$ was calculated using $\sigma_{xy}^A = \rho_{yx}^A/(\rho_{xx}^2 + \rho_{yx}^2)$.

Complex impedance and nonlinear Hall effects were measured using lock-in detection (SR-830 and SR-860, Stanford Research Systems; LI5650, NF Corporation). A sinusoidal excitation current, $I = I_0 \sin \omega t$, was applied, and both the in-phase ($V_{\mathrm{in}}^{\omega}$) and out-of-phase ($V_{\mathrm{out}}^{\omega}$) voltage components were recorded using a standard four-terminal configuration. The imaginary part of the linear impedance was defined as $\mathrm{Im}\, Z^{\omega} = V_{\mathrm{in}}^{\omega}/I$. Longitudinal and Hall components of the complex impedance were obtained by symmetrization and antisymmetrization with respect to the magnetic field.

To subtract the background signal originating from extrinsic phase rotations in cables and instruments, measurements at $T$ = 30 K, well above the magnetic transition temperature $T_C$ = 15 K, were used as a reference (see Supplementary Fig. S4). The data presented in the main text represent these background-subtracted values. The sample temperature $T_s$ was determined from the longitudinal resistivity as a thermometer, as described in Supplementary Fig. S2.

**Quantitative estimation of magnetization tilt angle**

We estimated the magnetization tilt angle induced by a high-density electric current by performing a control experiment with a deliberately tilted external magnetic field and comparing the resulting changes in the anomalous Hall effect between the two cases. The detailed procedure is described step by step below.

**Step 1**

Firstly, we calculate the anisotropic magnetic field $H_K$, which is the typical magnetic field required to tilt the magnetization away from the easy axis. Extended Data Fig. 1**a** shows the difference in magnetization, $\Delta M = M_{\parallel z} - M_{\perp z}$,where $M_{\parallel z}$ and $M_{\perp z}$ denote the magnetization with $H$ parallel and perpendicular to the $z$-axis, respectively. The anisotropy field $H_{\mathrm{K}}$ is defined by half the value of $H$ where $\Delta M = 0$. By fitting $\Delta M$ with a linear function, the value of $H$ at which $\Delta M = 0$ was determined to be $28\,\mathrm{T}$, resulting in $H_{\mathrm{K}} = 14\,\mathrm{T}$. The following model for the case of uniaxial magnetic anisotropy was used to estimate the tilt angle of magnetization from the $z$-axis ($\theta_{\mathrm{M}}$) induced by the external magnetic field.

$$E = -M_{\mathrm{S}} \frac{H_K}{2} \cos^2\theta_{\mathrm{M}} - \boldsymbol{M} \cdot \boldsymbol{H} \tag{5}$$

Here, $E$ is the energy of the spin system, $M_{\mathrm{S}}$ is the saturation magnetization, $\boldsymbol{M}$ is the magnetization, and $\boldsymbol{H}$ is the external magnetic field. The tilt angle $\theta_M$ was obtained by minimizing $E$. Specifically, $\theta_M$ was calculated by solving numerically the following equation:

$$\frac{\partial E}{\partial \theta_M} = -M_S[\sin(2\theta_{\mathrm{M}}) + H\sin(\theta_{\mathrm{M}} - \theta_{\mathrm{H}})] = 0 \tag{6}$$

Here, $H = |\,\boldsymbol{H}\,|$ and $\theta_{\mathrm{H}}$ is the tilt angle of the external magnetic field from the $z$-axis.

Extended Data Fig. 1**b** shows the result of the numerical calculation of Eq. (2) for $\theta_{\mathrm{H}} = 85°$.

**Step 2**

To compare magnetization tilt by high-density electric current and tilted external magnetic field, we here measured $\sigma_{xy}^{A}$ under two conditions: when the magnetic field was applied parallel to the $z$-axis and when it was tilted by $\theta_{\mathrm{H}} = 85°$ from the $z$-axis. In the tilted-field configuration, the in-plane component of the magnetic field was aligned parallel to the current direction. Ordinary Hall contribution was estimated as $R_0 B_z$, where $B_z$ is the $z$-axis component of the applied magnetic field and $R_0$ was obtained from the measurement under the field parallel to the $z$-axis. $\rho_{yx}^{A}$ was extracted by subtracting $R_0 B_z$ from the total $\rho_{yx}$ and $\sigma_{xy}^{A}$ was calculated using $\sigma_{xy}^{A} = \rho_{yx}^{A}/(\rho_{xx}^{2} + \rho_{yx}^{2})$. We then calculated the reduction of $\sigma_{xy}^{\mathrm{A}}$, denoted as $\Delta\sigma_{xy}^{\mathrm{A}}$, induced by tilted external magnetic field. The reduction was normalized by $\sigma_{xy}^{A0}$, which is the anomalous Hall conductivity when the magnetic field is applied parallel to the $z$-axis. The normalized reduction, $\Delta\sigma_{xy,\mathrm{B}}^{\mathrm{A}}/\sigma_{xy}^{\mathrm{A0}}$, is defined as follows:

$$\frac{\Delta\sigma_{xy,\mathrm{B}}^{\mathrm{A}}}{\sigma_{xy}^{\mathrm{A0}}} = \frac{\sigma_{xy}^{\mathrm{A}}\big(B = (B_x, 0, B_z)\big) - \sigma_{xy}^{\mathrm{A}}\big(B = (0,0,B_z)\big)}{\sigma_{xy}^{\mathrm{A}}\big(B = (0,0,B_z)\big)} \tag{7}$$

Here, the applied current density ($3 \times 10^{7}\ \mathrm{A/m^2}$) is sufficiently low so as not to induce the magnetization dynamics. Extended Data Fig. 1**c** shows the out-of-plane magnetic-field ($B_z$) dependence of the anomalous Hall conductivity for the external magnetic field $\theta_{\mathrm{H}} = 85°$ (red curve) and $\theta_{\mathrm{H}} = 0°$ (black curve).

**Step 3**

We plotted the normalized reduction of anomalous Hall conductivity, $\Delta\sigma_{xy,\mathrm{B}}^{\mathrm{A}}/\sigma_{xy}^{\mathrm{A0}}$, as a function of the magnetization tilt angle $\theta_M$ (Extended Data Fig. 1**d**). Note that the shaded area corresponds to the region where the magnetization is not saturated and is excluded for the current analysis. Remarkably, $\Delta\sigma_{xy,\mathrm{B}}^{\mathrm{A}}/\sigma_{xy}^{\mathrm{A0}}$ is smaller than $1-\cos\theta_{\mathrm{M}}$, which is a theoretical curve generally expected for the perturbative case where $\sigma_{xy}^{\mathrm{A}} \propto M_z$ (dashed line). This can be attributed to the change in electronic structure itself by magnetization tilting as discussed in Fig.2 in the main text.

**Step 4**

We next calculate the normalized reduction of anomalous Hall conductivity induced by electric current when the magnetic field is applied parallel to the $z$-axis. This normalized reduction, $\Delta\sigma_{xy,\mathrm{j}}^{\mathrm{A}}/\sigma_{xy}^{\mathrm{A0}}$, is defined as

$$\frac{\Delta\sigma_{xy,\mathrm{j}}^{\mathrm{A}}}{\sigma_{xy}^{\mathrm{A0}}} = \frac{\sigma_{xy}^{\mathrm{A}}(B=(0,0,B_z),j) - \sigma_{xy}^{\mathrm{A}}(B=(0,0,B_z),j\to 0)}{\sigma_{xy}^{\mathrm{A}}(B=(0,0,B_z),j\to 0)} \tag{8}$$

Here, the Hall conductivity measured under a current density of $3\times 10^{7}\ \mathrm{A/m^2}$ was used as $\sigma_{xy}^{\mathrm{A}}(B=(0,0,B_z),j\to 0)$, since this current density is sufficiently low and does not induce the magnetization tilt. Notably, $\Delta\sigma_{xy,\mathrm{j}}^{\mathrm{A}}/\sigma_{xy}^{\mathrm{A0}}$ takes a positive value at $j = 1\times 10^{8}\ \mathrm{A/m^2}$ and $2\times 10^{8}\ \mathrm{A/m^2}$. This behavior is attributed to temperature calibration errors, because in principle the application of electric current should reduce $\Delta\sigma_{xy,\mathrm{j}}^{\mathrm{A}}/\sigma_{xy}^{\mathrm{A0}}$. We estimate the uncertainty in $\Delta\sigma_{xy,\mathrm{j}}^{\mathrm{A}}/\sigma_{xy}^{\mathrm{A0}}$ to be within $\pm 1\%$, as the maximum value observed at $j = 1\times 10^{8}\ \mathrm{A/m^2}$ is approximately 1%. Extended Data Fig. 1**e** shows the magnetic field dependence of the normalized reduction of anomalous Hall conductivity by electric current, $\Delta\sigma_{xy,\mathrm{j}}^{\mathrm{A}}/\sigma_{xy}^{\mathrm{A0}}$.

**Step 5**

Assuming that the tilt angle of magnetization is determined solely by the normalized reduction of the anomalous Hall conductivity, the tilt angle induced by the electric current can be estimated by comparing with the result obtained under an external magnetic field. We assumed that when $\Delta\sigma_{xy,\mathrm{B}}^{\mathrm{A}}/\sigma_{xy}^{\mathrm{A0}} = \Delta\sigma_{xy,\mathrm{j}}^{\mathrm{A}}/\sigma_{xy}^{\mathrm{A0}}$ the tilt angle of magnetization $\theta_M$ is identical in both cases. For instance, the red horizontal line presented in Extended Data Fig.1**d** represents $\Delta\sigma_{xy,\mathrm{j}}^{\mathrm{A}}/\sigma_{xy}^{\mathrm{A0}}$ for $B_z = 0.4\,T$, $j = 6 \times 10^8\ \mathrm{A/m^2}$. This value equals to $\Delta\sigma_{xy,\mathrm{B}}^{\mathrm{A}}/\sigma_{xy}^{\mathrm{A0}}$ at $\theta_M = 8.8°$. Therefore, the tilt angle of the magnetization under $B_z = 0.4\,T$ $j = 6 \times 10^8\ \mathrm{A/m^2}$ is estimated to be $\theta_M = 8.8°$.

**Step 6**

By repeating the above analysis, the current-density dependence of $\theta_M$ was obtained (Fig. 2**c**). The error in $\theta_M$ was estimated as

$$\delta\theta_M = \frac{d\theta_M}{d\left(\frac{\Delta\sigma_{xy,j}^{A}}{\sigma_{xy}^{A0}}\right)} \times \delta\left(\frac{\Delta\sigma_{xy,j}^{A}}{\sigma_{xy}^{A0}}\right) \tag{9}$$

Here, $\delta(\Delta\sigma_{xy,j}^{A}/\sigma_{xy}^{A0})$ was assumed to be 1% as described in step 4, and $d\theta_M/d(\Delta\sigma_{xy,j}^{A}/\sigma_{xy}^{A0})$ was approximated from the slope at each point in the $\theta_M$ versus $\Delta\sigma_{xy,j}^{A}/\sigma_{xy}^{A0}$ plot (Extended Data Fig.1**f**).

**DFT calculation**

The electronic-structure calculations for PrAlGe were performed using density functional theory (DFT) as implemented in the Vienna Ab initio Simulation Package (VASP)[34,35]. The plane-wave kinetic-energy cutoff was set to 500 eV to properly describe systems containing rare-earth atoms. Projector augmented-wave (PAW) pseudopotentials[36] were used together with the generalized gradient approximation (GGA) in the Perdew–Burke–Ernzerhof (PBE) form Refs. [37,38]. The Brillouin zone was sampled with a Γ-centered Monkhorst–Pack mesh of 10 × 10 × 10 k-points. The PAW valence configurations were: Ge ($4s^2$ $4p^2$), Al ($3s^2$ $3p^1$), and Pr ($5s^2$ $5p^6$ $4f^3$ $6s^2$). The crystal structure was taken from experiment[39]: space group $I4_1md$ (No. 109), a = b = 4.2642 Å, c = 14.685 Å. The atoms were placed at the 4a Wyckoff positions: Ge (0,0,0), Al (0,0,0.166), and Pr (0,0,0.581). The localized Pr 4f-states were treated within the rotationally invariant DFT+U scheme[40] with a Hubbard U = 6 eV. The canted magnetic configuration (tilt of θ = 15°) was stabilized by applying constraining fields via VASP's penalty functional[41].

The projected Wannier functions were generated using the Wannier90 code[42] and the VASP2WANNIER90 interface; the maximal-localization procedure was not applied. The electronic band structures and Berry curvatures for 1D and 2D cuts were computed using the Wannier90 post-processing tools[42]. The 2D maps of the Berry curvature were generated on a 200 × 200 *k*-mesh in the (*xy*)-plane and integrated along $k_z$ using a 101 *k*-mesh. Lastly, the locations of the Weyl points were identified using the Wannier Tools software[43]. We employed a discretization of 15×15×15 *k*-points within the Brillouin zone and an energy-gap threshold of 0.010 eV. Using these settings, we identified 56 Weyl nodes for $\theta = 0°$ (magnetization along the *c*-axis) and 352 Weyl nodes for the canted configuration $\theta = 15°$ (magnetization in the *xz* plane).

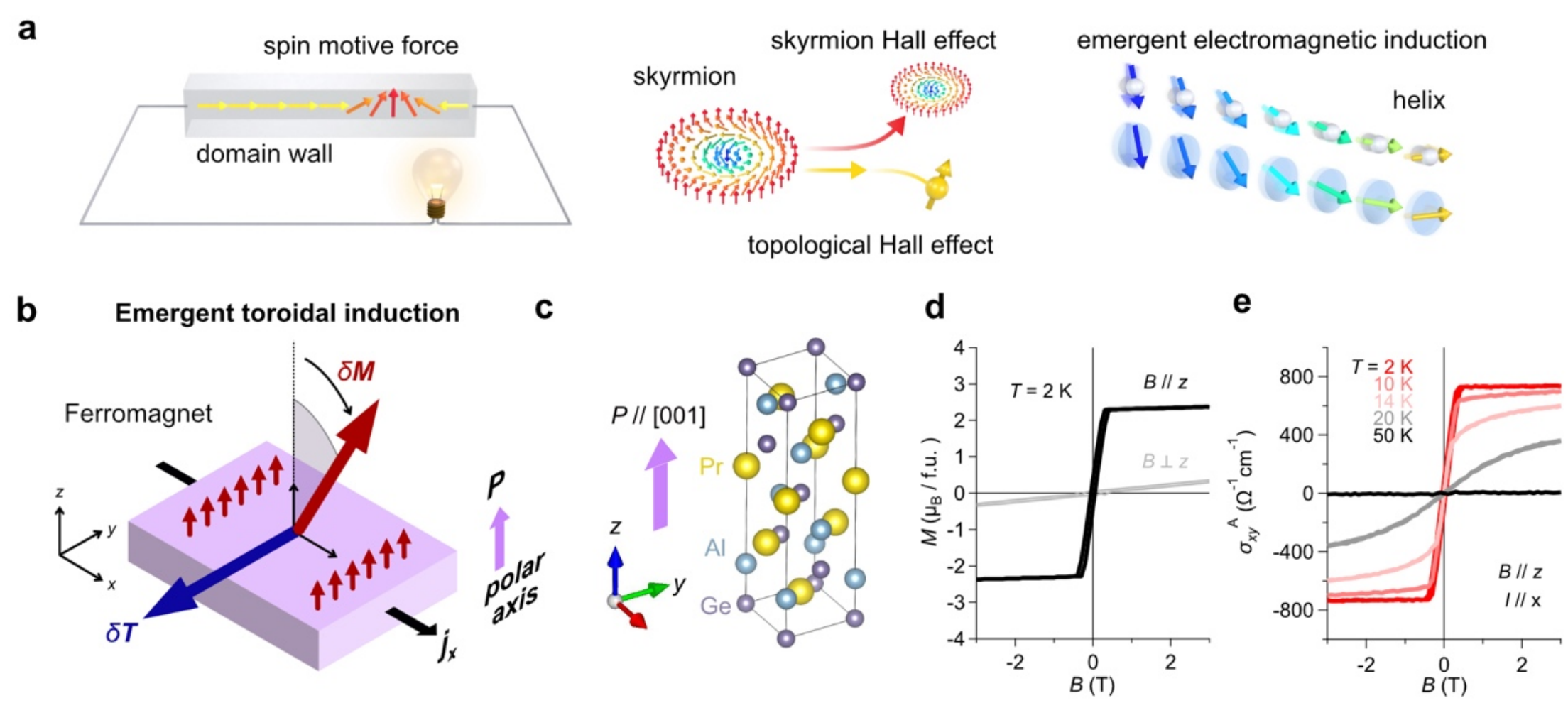

**Fig. 1 | Current-induced control of magnetization in the polar Weyl ferromagnet PrAlGe. a,** Schematics of emergent electric fields arising from *non-uniform* spin textures. From left to right: spin motive force generated by a moving domain wall, emergent electric field induced by skyrmion motion, and emergent electromagnetic induction in a current-driven helical structure. **b**, Schematics of current-induced magnetization tilt ($\delta \boldsymbol{M}$) in a polar ferromagnet and the corresponding variation of the toroidal moment ($\delta \boldsymbol{T}$). **c**, Crystal structure of PrAlGe. **d**, Magnetization curves at $T$ = 2 K for external magnetic fields applied parallel (black) and perpendicular (gray) to the polar axis ($z$-axis). **e**, Anomalous Hall effect measured at various temperatures.

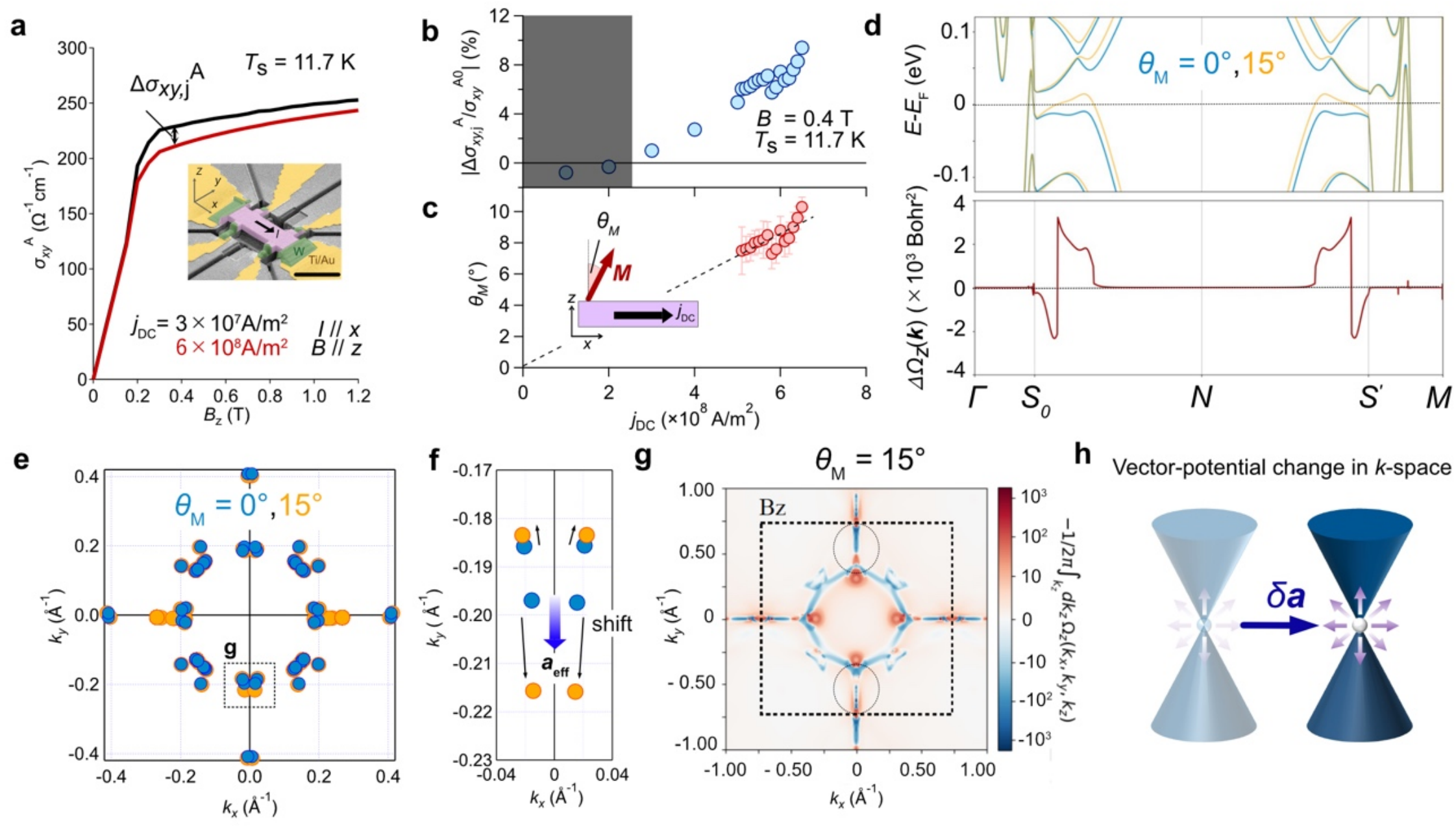

**Fig. 2 | Direct-current induced magnetization dynamics and their impact on the electronic band structure of PrAlGe. a**, Anomalous Hall conductivity (AHC) ($\sigma_{xy}^{\mathrm{A}}$) measured at low (black) and high (red) current densities, $j_{\mathrm{DC}} = 3 \times 10^7$ A/m$^2$ and $j_{\mathrm{DC}} = 6 \times 10^8$ A/m$^2$, respectively. The inset shows a scanning electron micrograph of the PrAlGe device with false colors. Scale bar, 10 µm. **b**, Normalized change in AHC at $B = 0.4$ T as a function of current density, defined as $\Delta\sigma_{xy}^{\mathrm{A}}/\sigma_{xy}^{\mathrm{A0}} \equiv [\sigma_{xy}^{\mathrm{A}}(j_{\mathrm{DC}}) - \sigma_{xy}^{\mathrm{A}}(j_{\mathrm{DC}} = 3 \times 10^7)]/\sigma_{xy}^{\mathrm{A}}(j_{\mathrm{DC}} = 3 \times 10^7)$. **c**, Estimated tilt angle ($\theta_M$) along the *x*-direction relative to the easy axis (z-axis) as a function of current density (see Extended Data Fig. 1 for detail). The sample temperature is kept at $T_s = 11.7$ K for panels **a**–**c**, independent of $j_{\mathrm{DC}}$. **d**, Calculated electronic band structure for $\theta_M = 0^{\circ}$ (blue) and $\theta_M = 15^{\circ}$ (yellow) (top), and the corresponding change in the momentum-resolved Berry curvature integrated in the $k_x$-$k_y$ plane $\Delta\Omega_z(\boldsymbol{k}) \equiv \Omega_z(\boldsymbol{k})^{\theta_M=15^{\circ}} - \Omega_z(\boldsymbol{k})^{\theta_M=0^{\circ}}$ (bottom). **e**, Calculated 2D map of integrated Berry curvature in the $k_x$-$k_y$

plane. The bold dashed square denotes the first Brillouin zone (BZ), and the thin dashed circles highlight regions with marked changes in Berry curvature. **f**, Locations of Weyl nodes in the $k_z = 0$ plane for $\theta_M = 0^{\circ}$ (blue) and $\theta_M = 15^{\circ}$ (yellow). **g**, Enlarged views of the dashed squares in **f**, where the shift of the Weyl nodes generates an effective vector potential along the $k_y$-direction. **h**, Schematic illustration of the change in vector potential ($\delta\boldsymbol{a}$) in momentum space due to a shift of the Weyl-node position.

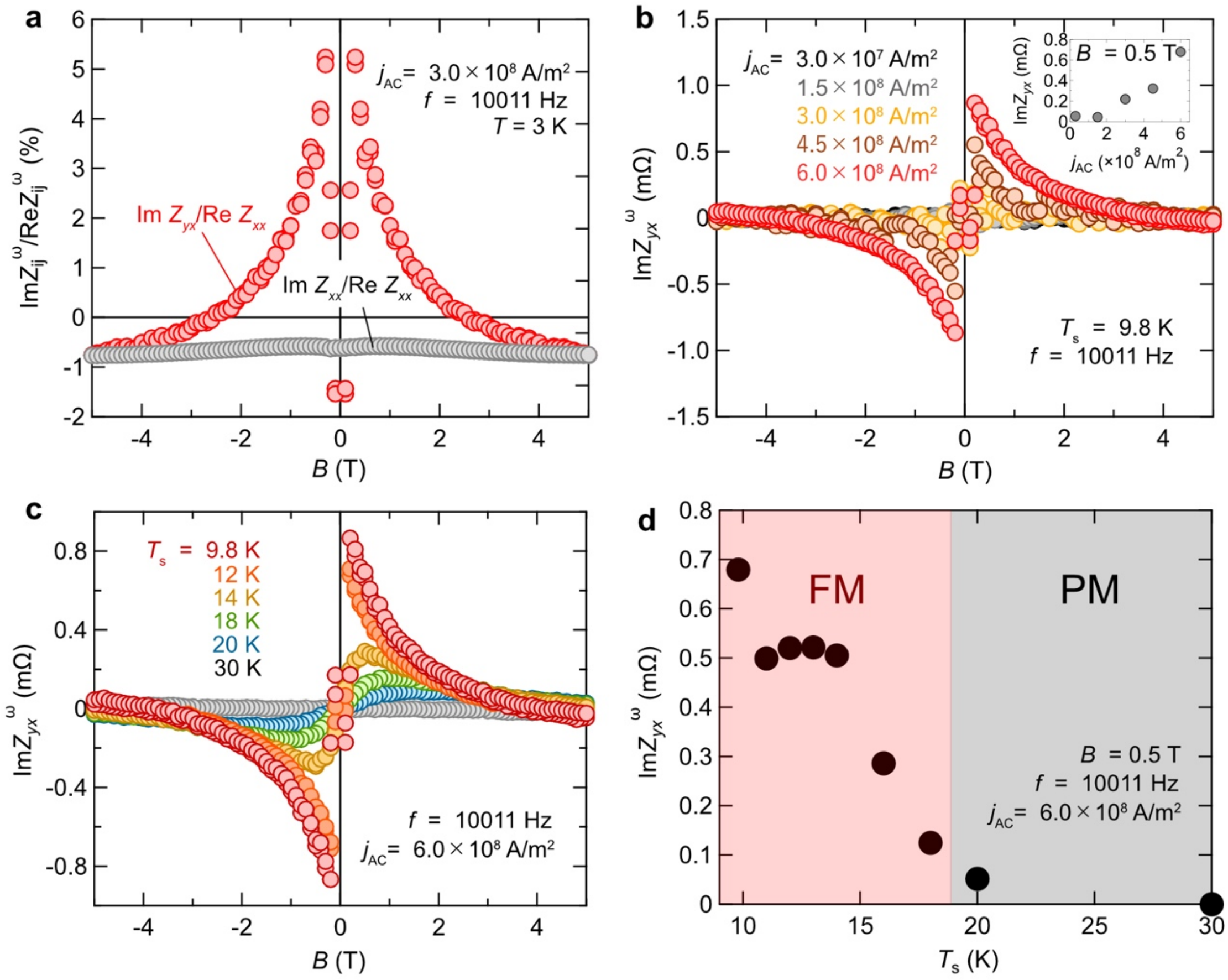

**Fig. 3 | Emergent electromagnetic Hall induction in PrAlGe. a**, Magnetic-field dependence of the imaginary part of the complex impedance ($\mathrm{Im}Z_{ij}^{\omega}$) normalized by its real part ($\mathrm{Re}Z_{ij}^{\omega}$), measured with $j_{\mathrm{AC}} = 3.0 \times 10^8$ A/m², $f = 10$ kHz, and $B$ // $c$-axis at $T = 3$ K. The red and gray dots represent the transverse (Hall) and longitudinal complex impedance, respectively. **b**, Magnetic-field dependence of the imaginary part of the complex Hall impedance for various current-densities measured at $f = 10011$ Hz and $B$ // $c$-axis at a sample temperature $T_{\mathrm{s}}$ = 9.8 K. The inset shows the current-density dependence of the imaginary part of the complex Hall impedance at $B$ = 0.5 T. **c**, Magnetic-field dependence of the imaginary part of the complex Hall impedance at various sample temperatures $T_{\mathrm{s}}$ measured with $j_{\mathrm{AC}} = 6.0 \times 10^8$ A/m², $f = 10011$ Hz, and $B$ // $c$-axis. **d**, Imaginary part of the complex Hall impedance at a fixed magnetic field

$B$ = 0.5 T as a function of sample temperature $T_s$. The red and gray regions denote the ferromagnetic (FM) and paramagnetic (PM) phases, respectively.

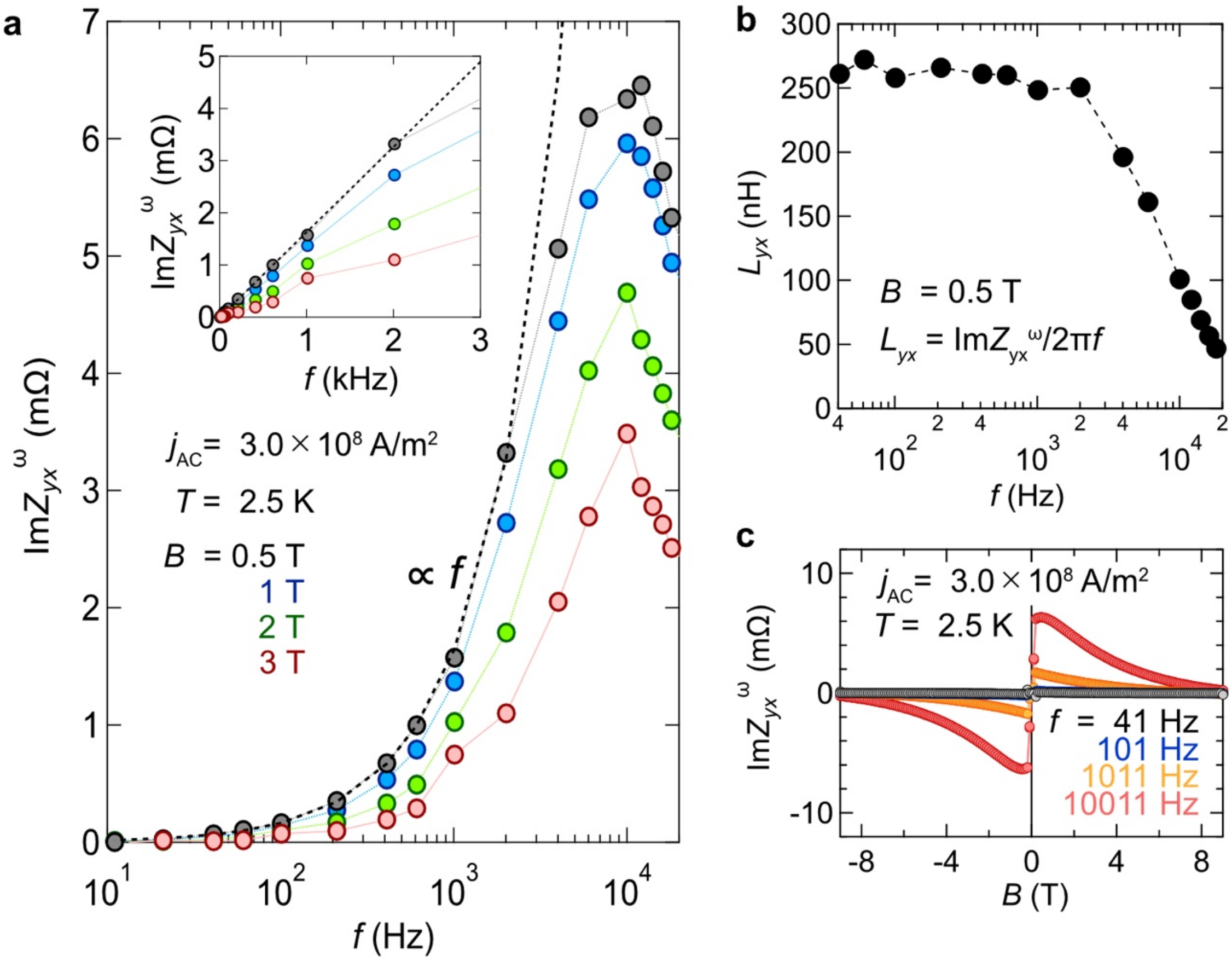

**Fig. 4 | Frequency dependence of emergent electromagnetic Hall induction in PrAlGe. a**, Frequency dependence of the imaginary part of the complex Hall impedance (Im $Z_{yx}^{\omega}$) measured with $j_{\mathrm{AC}} = 3.0 \times 10^8$ A/m$^2$ and $B$ // $c$-axis at various magnetic fields, showing a linear frequency dependence as indicated by the dashed black curve. The inset shows the linear-linear plot for the regime where the complex Hall impedance is proportional to frequency. **b**, Temperature dependence of the real part of the complex Hall inductance $L_{yx} = ImZ_{yx}^{\omega}/2\pi f$ at $B$ = 0.5 T. **c**, Magnetic-field dependence of the imaginary part of the complex Hall impedance at $T$ = 2.5 K for various frequencies, measured with $j_{\mathrm{AC}} = 3.0 \times 10^8$ A/m$^2$.

## Acknowledgements

This work was supported by the JSPS KAKENHI Grant Number 23H05431, 24H00197, 24H02231, 24K00583, 24K16986 and JST PRESTO JPMJPR2597, JPMJPR23HA. N.N. was supported by the RIKEN TRIP initiative. The authors are grateful to C. Terakura and M. Ishida for preparing schematic illustrations.

## Author Contributions

FIB microdevice was fabricated by Y.S. and Y.F. with the support from M.T.B.. A.K. synthesized the PrAlGe single crystal under the supervision of Y. Ta. Transport and lock-in measurements were performed by Y.S., Y.F., and M.M. with the assistance from D.Y., M.T.B., Y.K., M.K.. The DFT calculations were conducted by J.B. and R.A.. Theoretical models were discussed with T.A., T.N., and N.N.. Manuscript was written by Y.F., M.M., and Y.S. with the support from J.B., T.A., A.K., M.T.B., M.K., Y. Ta., T.M., R.A., N.N., and Y.To.. All the authors commented on the manuscript. The project was conceived and supervised by Y.F. and Y. To.

## Competing Interests

The authors declare no competing interests.

## Materials & Correspondence

Correspondence and requests for materials should be addressed to Y.F. and Y.T..

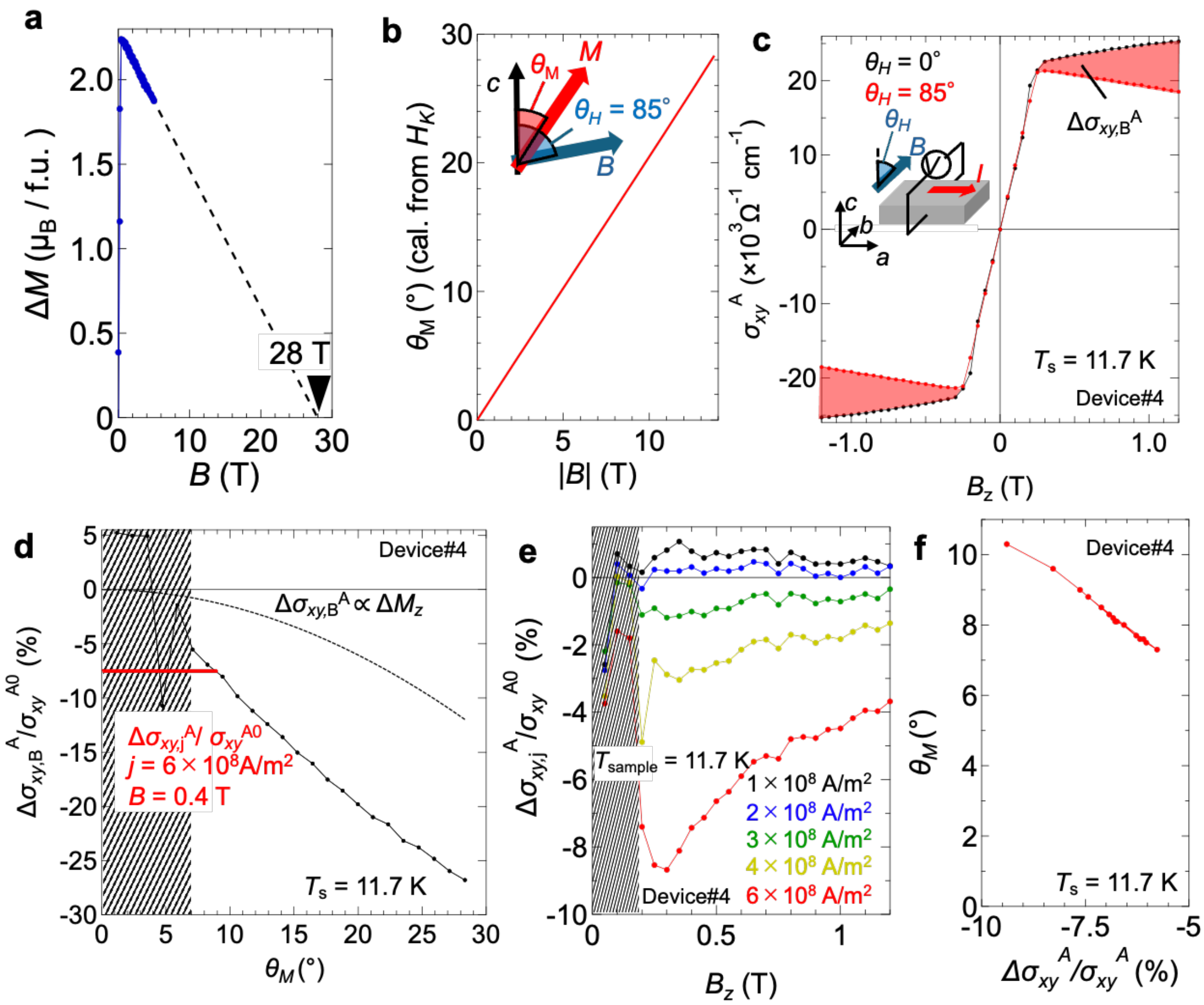

**Extended Data Fig. 1 | Quantitative estimation of magnetization tilt angle. a,** The difference in magnetization ($\Delta M = M_{\parallel z} - M_{\perp z}$) as a function of magnetic field. Here, $\Delta M = M_{\parallel z} - M_{\perp z}$, where $M_{\parallel z}$ and $M_{\perp z}$ denote the magnetization with $H$ parallel and perpendicular to the $z$-axis, respectively. **b,** Calculated tilt angle ($\theta_M$) as a function of external magnetic field, which is applied with the tilted angle of $\theta_H = 85^{\circ}$ from the z-axis. The inset shows the schematic illustration of the experimental setup and the definition of $\theta_M$ and $\theta_H$. **c,** Anomalous Hall conductivity (AHC) as a function of $z$-component of the external magnetic field. The black(red) curve denotes the case where the external magnetic field is applied with $\theta_H = 0^{\circ}$ ($\theta_H = 85^{\circ}$). The difference between these two data is defined as $\Delta\sigma_{xy,\mathrm{B}}^{\mathrm{A}}$. **d,** Normalized reduction of AHC induced by tilted

external magnetic field, plotted against $\theta_M$ (see Methods for detailed definition). The shaded area corresponds to the region where the magnetization is not saturated and is excluded for the current analysis. The dashed curve represents $1 - \cos\theta_{\mathrm{M}}$, which is a theoretical curve generally expected for the case when $\sigma_{xy}^{\mathrm{A}} \propto M_z$. **e,** Normalized reduction of AHC induced by electric current when the magnetic field is applied parallel to the $z$-axis, plotted against the external magnetic field (see Methods for detailed definition). **f,** $\theta_{\mathrm{M}}$ plotted as a function of normalized reduction of AHC induced by electric current.

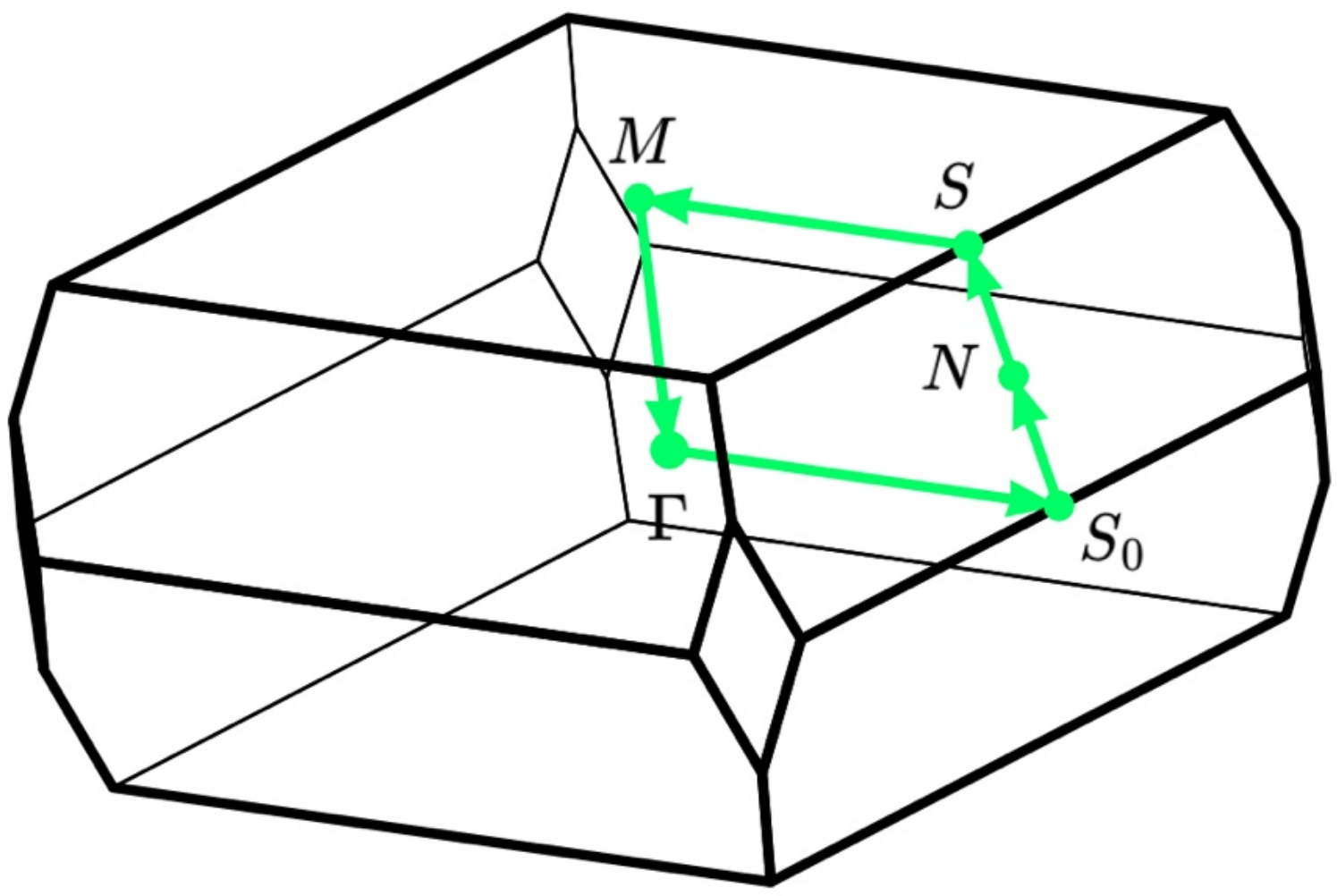

**Extended Data Fig. 2 | Definition of the first Brillouin zone and the k-path (green line) for the Fig. 2d.**